\documentclass[a4paper,fleqn,usenatbib]{mnras}

\usepackage{newtxtext,newtxmath}

\usepackage[T1]{fontenc}
\usepackage{ae,aecompl}

\usepackage{graphicx}	
\usepackage{amsmath}	
\usepackage{amssymb}	

\title[Electron acceleration at sub- and supercritical shocks]{Electron acceleration at quasi-perpendicular shocks in sub- and supercritical regimes: 2D and 3D simulations}

\author[Trotta \& Burgess]{
D. Trotta $^{1}$\thanks{E-mail: d.trotta@qmul.ac.uk}
and D. Burgess$^{1}$
\\
$^{1}$Queen Mary University of London, School of Physics and Astronomy, London E1 4NS, UK
}

\date{Accepted XXX. Received YYY; in original form ZZZ}

\pubyear{2018}

\begin{document}
\label{firstpage}
\pagerange{\pageref{firstpage}--\pageref{lastpage}}
 \newcommand{\quomark}[1]{``#1''}

\maketitle

\begin{abstract}

Shock accelerated electrons are found in many astrophysical environments, and the mechanisms by which they are accelerated to high energies are still not completely clear. For relatively high Mach numbers, the shock is supercritical, and its front exhibit broadband fluctuations, or ripples. Shock surface fluctuations have been object of many observational and theoretical studies, and are known to be important for electron acceleration. We employ a combination of  hybrid Particle-In-Cell and test-particle methods to study how shock surface fluctuations influence the acceleration of suprathermal electrons in fully three dimensional simulations, and we give a complete comparison for the 2D and 3D cases. A range of different quasi-perpendicular shocks in 2D and 3D is examined, over a range of parameters compatible with the ones observed in the solar wind. Initial electron velocity distributions are taken as kappa functions, consistent with solar wind \emph{in-situ} measurements.
Electron acceleration is found to be enhanced in the supercritical regime compared to subcritical. When the fully three-dimensional structure of the shock front is resolved, slightly larger energisation for the electrons is observed, and we suggest that this is due to the possibility for the electrons to interact with more than one surface fluctuation per interaction. In the supecritical regime, efficient electron energisation is found also at shock geometries departing from $\theta_{Bn}$ very close to 90$^\circ$. Two dimensional simulations show indications of unrealistic electron trapping, leading to slightly higher energisation in the subcritical cases.

\end{abstract}




\section{Introduction}

Electron acceleration at collisionless shocks is a key process in space and astrophysical plasmas, being observed in situ at planetary bow shocks \citep[e.g.][]{Burgess2007_rev,Masters2017}, and interplanetary shocks \citep[e.g.][]{Potter1981,Dresing2016}. It is also inferred from observations of solar radio Type II emission \citep[e.g.][]{Holman1983,Pulupa2008}, synchrotron emission at SNR shocks \citep[e.g.][]{Koyama1995,Ellison2001} and diffuse radio emission from the intragalactic cluster medium \citep[e.g.][]{Ensslin1998,Kang2017}. 

Shocks in general convert directed flow energy (upstream) to thermal energy (downstream), and at shocks in collisionless plasmas a small fraction of the energy is available for acceleration of particles to high energies.
Space observations and simulation studies have shown that the internal structure of the shock is important for the relevant type of acceleration mechanism, and also its detailed operation \citep{burgess_book}. Depending on its Mach number the shock can be sub- or supercritical, where, for the latter, ion reflection and gyration dominates and controls both the average shock structure and the types of microstructure associated with instabilities. Recent three-dimensional hybrid simulations (kinetic ions and fluid electrons) have revealed more detail of the microstructure of quasi-perpendicular supercritical shocks \citep{Burgess2016}. In this paper we will explore the effects of this microstructure on electron acceleration, starting from just above thermal energies. We will also demonstrate the importance of the sub- and supercritical Mach number regimes for the effectiveness of shocks as sources of electron acceleration; this is important when invoking shocks as electron acceleration sites for any particular astrophysical system.

Collisionless shock transitions have an internal structure controlled by many parameters,the most important of which is the angle between the upstream magnetic field and the normal to the shock surface, $\theta_{Bn}$. When $\theta_{Bn}$ $\gtrsim$ 45$^\circ$ (i.e., the upstream magnetic field is almost parallel to the shock surface), the shock is quasi-perpendicular; and when $\theta_{Bn}$ $\lesssim$ 45$^\circ$, the shock is quasi-parallel. In this work we concentrate on electron acceleration at quasi-perpendicular shocks, motivated by both observational and theoretical arguments. 

Shock accelerated electrons were first observed in situ upstream of the Earth's bow shock, in the region known as the electron foreshock \citep{Anderson1969}.  Energetic electrons were observed with energies 50 eV to $>$10 keV streaming away from the bow shock, along magnetic field lines connected to the shock surface. The most energetic backstreaming electrons were found to be confined in a small region, corresponding immediately downstream of the tangent point between the interplanetary magnetic field and the bow shock surface (where $\theta_{Bn} = 90^\circ$). Furthermore, the source was found to be ordered in energy, with less energetic electrons found deeper (further downstream) into the electron foreshock \citep{Gosling1989}. The electrons at intermediate energies (up to about 1 keV) originate in a broad region behind the magnetic tangent surface, i.e., on field lines with connection to the shock at $\theta_{Bn} < 90^\circ$. A thorough review of these observations can be found in \citet{fitz1995}.

The first analytical model for accelerated electrons at quasi-perpendicular shocks was based on adiabatic reflection \citep{Leroy1984,Wu1984}, which assumes a 1D, planar and steady shock, and magnetic moment conservation in the Hoffman - de Teller frame (HTF) resulting in magnetic mirror reflection for some electrons. The HTF is the shock frame in which the flow is parallel to the magnetic field, and the motional electric field is zero (assuming ideal MHD, although it can be nonzero within the shock transition). In the HTF, the electron energy is constant, neglecting any change due to electric field in the shock structure.  Here the energy gain is associated with the frame transformation from the HTF to the observer frame. In the observer frame (typically the Normal Incidence Frame, NIF, where upstream bulk flow and shock normal are parallel), the electrons gain energy in the reflection process via drift parallel to the motional electric field. The energy gain for reflected electrons depends on the velocity frame transformation from the NIF to the HTF, so increases with $\theta_{Bn}$, and for significant energisation, $\theta_{Bn}$  has to be close to 90$^\circ$. On the other hand, the density of reflected electrons decreases as $\theta_{Bn}$ increases since only electrons from the wings of the incident distribution satisfy the conditions for reflection. The presence of a cross-shock potential modifies the reflection process, acting to reduce reflection at low energies. The  resulting distribution function of reflected electrons is a truncated loss-cone \citep{Leroy1984}.

The role of self consistent shock structure (e.g., overshoot, cross-shock potential) have been investigated by means of test particle simulation of electrons using electromagnetic fields obtained from 1D plasma simulations \citep{Krauss1989,Krauss1989b}. Later, the role of shock curvature was found to be important in terms of electron energisation, due to the relatively large distance transverse to the shock travelled by electrons during reflection \citep{Krauss1991}. Other studies \citep[e.g.,][]{Zlobec1993,Vandas2002,Knock2003} linked the presence of large scale ripples at the shock surface features to electron acceleration in interpreting solar type II radio bursts.

The shock structure depends crucially on the Mach number. A shock can be sub- or supercritical, depending on its Mach number relative to the critical Mach number $M_c$, defined as that at which the downstream flow speed is equal to the speed of sound. Super-critical shocks require a dissipation process other than resistivity, and this is provided by ion reflection and gyration into the downstream. Although the definition of $M_c$ arises from two-fluid theory, usually supercritical shocks are treated as those dominated, in terms of structure and thermalization, by ion reflection and gyration. The structure is usually described as "foot-ramp-overshoot" where the foot is formed by reflected ions gyrating around ahead of the main ramp before returning to the shock. On the other hand low Mach number, subcritical shocks do not exhibit strong structuring around the transition layer, appearing similar to laminar fluid shocks.

Non-stationarity and microstructure are important features of collisionless shocks. Self-reformation of supercritical, quasi-perpendicular shocks, with a quasi-periodic steepening of the shock ramp, has been found in simulations \citep[e.g.,][]
{Biskamp1972,Quest1985,Lembege1992}. The process of self-reformation is important at low $\beta_i$ (i.e., the ratio of upstream ion plasma to magnetic field pressures), and high Mach number \citep{Hada2003}. It can be inhibited through the emission of nonlinear whistler waves in the shock foot \citep{Krasno1991,Hellinger2007}, although the situation can be complicated by the shock geometry used in the simulations \citep{Lembege2009}. In addition, the shock ramp and foot can be unstable to multiple wave modes, leading to microstructure within the shock transition. The landscape of possible microinstabilities that can be generated in the foot of quasi-perpendicular shocks is broad: \citet{Matsukiyo2006} identified six types of instabilities being excited in short times (less than one ion gyroperiod) using 2D, fully kinetic simulations. Early hybrid 2D simulations of perpendicular shocks with the magnetic field in the simulation plane showed that the surface exhibits fluctuations or ripples associated with ion Alfv\'en cyclotron and possibly mirror instabilities \citep{Winske1988}. The ripples propagate across the shock front at the Alfv\'en speed of the shock overshoot \citep{Lowe2003}. Simulations of perpendicular shocks with the magnetic field out of the simulation plane revealed a different type of fluctuation, directly connected with the reflected ion population \citep{Burgess2007}. Recently, a study of structuring in 3D shock simulations has shown that 2D simulations do not fully capture the dynamics of shock structure since there are processes due to the coupling between field parallel and reflected ion fluctuations \citep{Burgess2016}. Another source of nonstationarity which has been identified at quasi-perpendicular shocks is whistler wave turbulence \citep[e.g.,][]{Krasno1991}. \citet{Oka2006}, using Geotail data, related the electron acceleration at the Earth's bow shock with the presence of whistler waves in the shock foot. Recent observational results obtained using the Magnetospheric Multiscale Spacecraft (MMS) have proven directly and for the first time that quasi-perpendicular, collisionless shocks do have a rippled surface \citep{Johlander2016}. It has been shown that the observed ripples are consistent with results from hybrid simulations, and are modulated by the process of shock reformation \citep[][]{gingell2017}. An extensive review about the dynamics of quasi-perpendicular shocks and their observational properties can be found in \citet{Krasno2013}.

A study of the role of surface ripples in electron acceleration was carried out by \cite{Burgess2006}, by means of test particle and 2D hybrid shock simulation. At low Mach numbers a good agreement with adiabatic reflection theory was found, whereas at high Mach numbers the rippled character of the shock enhanced the electron acceleration. This picture of enhanced energisation was extended by considering a turbulent upstream plasma flow, and it was found that electron energy gains are enhanced even at more oblique configurations (departing from the condition for $\theta_{Bn}$ to be close to 90$^\circ$) \citep{Guo2010}. This is due to the fact that large scale upstream fluctuations can mirror the electrons back to the shock, creating a multiple-shock encounter scenario which leads to larger electron energisation \citep{Guo2015}.
 
In the studies discussed so far, the shock structure has been assumed to be dominated by ion scale processes, making the hybrid simulation method appropriate. There are also studies of electron acceleration using fully kinetic Particle-In-Cell (PIC) simulations, which also model electron scale physics. Simulations which display nonstationarity (shock reformation) were found to eject upstream electrons in a bursty fashion \citep{Lembege2002}. Other simulations emphasized the importance of  whistler waves excited in the foot of the quasi-perpendicular shock for electron acceleration out of the thermal population, with relevance to super nova remnant shocks \citep{Riquelme2011}. Strong electron acceleration was also observed in full PIC simulations of high Mach number shocks by \citet{Amano2009}, where it was found that electrons were efficiently reflected at the shock front by electron-scale electrostatic fluctuations induced by the modified two-stream instability, and then accelerated by the convective electric field in front of the shock, in a picture known as ``electron surfing acceleration''. At lower Mach numbers, electron surfing acceleration was found to be important for both energisation itself and also as a channel of pre-acceleration for further energisation by shock drift acceleration \citep{Amano2007}. However, fully kinetic simulations have some limitations, for example, using a reduced proton to electron mass ratio can produce large differences in the shock structure compared to when the real ratio is used \citep[e.g.,][]{Scholer2003} (see also \citet{Krasno2013} for a discussion).
 
The motivation of this work is to extend earlier studies, in the picture of hybrid and test-particle simulations, comparing 2D and 3D simulations for a range of non-reforming shock parameters, and using realistic upstream kappa electron distributions, for which there is observational evidence in the solar wind \citep[e.g.,][]{Pierrard1999}. In this framework, we want to demonstrate the importance of the critical Mach number for electron acceleration, being an important issue for many astrophysical environments.
 
The paper is organised as follows: in section \ref{sec:method} the details of the hybrid and test-particle method are presented; section \ref{sec:shock_fluc} shows the different types of surface fluctuations found in the sub- and supercritical regimes; in section \ref{sec:2D_3D} the upstream energy spectra obtained in 2D and 3D simulations are compared; in section \ref{sec:3D_spectra} we focus on electron acceleration in 3D simulations, showing the scenarios for different shock parameters;  in section \ref{sec:discussion} we discuss the electron acceleration mechanisms in the simulations and in section \ref{sec:conclusions} the results are summarised.
  
\section{Method}
\label{sec:method}
We use and extend a method seen in earlier works \citep[e.g.,][]{Krauss1989, Burgess2006}, which consists of a combination of hybrid plasma and electron test particle simulations. In the hybrid plasma simulation method, the electrons are modelled as a massless, charge 
neutralising fluid with an adiabatic equation of state.  The protons are modelled as macroparticles and advanced using standard PIC methods. The hybrid code used (HYPSI), is based on the Current Advance Method - Cyclic Leapfrog (CAM-CL) algorithm \citep[][]{Matthews1994}. The electrons participating in acceleration are modelled as test particles in the time varying electric and magnetic fields from the hybrid simulation. This model assumes that the test particle electrons do not belong to the thermal population of the plasma, which is already modelled as a fluid in the hybrid framework. In other words, the initial energy for the test-particle electrons must be high enough (i.e., suprathermal) so that the feedback of the test particle ensemble to the macrospcopic fields can be neglected.

The simulation is carried out in two phases. First, a hybrid shock simulation is performed to obtain electric and magnetic fields with data stored for every grid point at every time step. The equations of motion for an ensemble of test particle electrons are then integrated using the time-dependent fields {using} interpolation in space and time at the particle position. The shock is initiated by the injection method, in which the plasma flows along the $x$-direction with a velocity $V_i$. The right-hand boundary of the simulation domain acts as a reflecting wall, and at the left-hand boundary plasma is continuously injected. The simulation is periodic in the $y$ and $z$ directions. A shock is created as a consequence of reflection at the wall, and it propagates in the negative $x$-direction. In the simulation frame, the upstream flow is along the  shock normal.

Distances are normalised to the ion inertial length $c/\omega_{pi} \equiv d_i$, time to the inverse cyclotron frequency ${\Omega_{ci}}^{-1}$, velocity to the Alfv\'en speed $v_A$ (all referred to the upstream state), and the magnetic field and density to their upstream values, $B_0$ and $n_0$, respectively. Results are presented for hybrid simulations in 2D and 3D, for a range of inflow velocity (and hence shock Mach numbers), and angle between the upstream magnetic field and the $x$-axis $\theta_{Bx}$. The latter corresponds to the angle between the upstream magnetic field and the mean shock normal, $\theta_{Bn}$. In all the simulations, the upstream magnetic field is in the $x$-$y$ plane. The simulation domain  is 60 $\times$ 60 $d_i$ for the 2D case, and 60 $\times$ 60 $\times$ 20 $d_i$ in the 3D case. The spatial resolution used is $\Delta x$ = $\Delta y$ = $\Delta z$ = 0.25 $d_i$. The time step for particle {(ion)} advance is $\Delta t_{pa}$  = 0.01 $\Omega_{ci}$. Substepping is used for the magnetic field advance, with an effective time step of $\Delta t_{B} = \Delta t_{pa}/10$. A small, nonzero  resistivity is introduced in the magnetic induction equation. The value of the resistivity is chosen so that there are not excessive fluctuations at the grid scale, and the overall shock behaviour is similar to that observed at the Earth's bow shock. For the parameters chosen here the average shock structure is quasi-stationary (i.e., not exhibiting self-reformation). The number of particles per cell used is always greater than 100 (upstream), in order to keep the statistical noise characteristic of PIC simulations to a reasonable level. 

Using the electromagnetic fields from the hybrid simulations, the equations of motion for an ensemble of  test-particle electrons are solved using a 4th order scheme \citep{Thomson1968}.  The electric and magnetic fields at each electron position are interpolated from the hybrid simulation grid. It is important, as shown in previous works \citep[][]{Burgess2006}, to employ a smooth spatial interpolation method to avoid artificial energisation. In these simulations three-dimensional spatial interpolation is carried out by means of a \quomark{tricubic} routine, which gives an interpolant function for the electromagnetic fields that is both $C^1$ and isotropic. It is also worth noting that this interpolation method is completely local, i.e., the interpolated function values depend only on the values at neighbouring grid points \citep{Lekien2005}. The interpolation in time, on the other hand, is linear. When advancing the electrons adaptive time stepping is used with a time step varying between upper and lower limits of 0.1 and 6 $\times$ 10$^{-6}$ ${\Omega_{ce}}^{-1}$.

The test particle electrons are released as a set of monoenergetic shells, with initial positions in a plane parallel to  the shock front and 5 $d_i$ upstream of it. The shock position is calculated at each time step using an average shock velocity, which in turn is based on the position at which the magnetic field first exceeds twice the upstream magnetic field. The time of release is chosen when the shock is already well developed, with the shock at least 15 $d_i$ away from the right-hand wall. The electrons are collected after the interaction with the shock at two boundaries placed respectively at 7 and 5 $d_i$ upstream and downstream, respectively, of the shock front. The energy shells are defined in eV, meaning that the free parameter $c/v_{A}$ is introduced when converting the energy value to the normalised units used by the hybrid simulation. We use a $c/v_A$ ratio of 5000, corresponding to an Alfv\'en speed of 60 km s$^{-1}$, typical of the solar wind. The velocity distribution for each energy shell is a sphere in velocity space, centred on the upstream inflow velocity. The final upstream and downstream distributions are reconstructed from the results of many initial energy shells. In this work the initial upstream electrons have a kappa distribution, as widely observed in the solar wind \citep[e.g.,][]{Maksimovic2005,Stevrak2009}. The test particle upstream kappa distribution has the following form:
\begin{eqnarray}
\label{eq1}
f_k(v)& =  & \frac{1}{(\pi \omega_k^2)^{3/2}} \, \frac{\Gamma(k+1)}{\Gamma(k-1/2)}\left( 1 + \frac{v^2}{k \omega_k^2}\right)^{-(k+1)} ,\\
\omega_k^2 &= & \left(1-\frac{3}{2k}\right) \left(\frac{2k_B T}{m}\right) ,
\end{eqnarray}
where $m$ is the electron mass, $k_B$ is the Boltzmann constant, $T$ is the electron 
temperature and $\Gamma$ denotes the Euler gamma function. For large $k$ the distribution approaches a Maxwellian. Throughout this study $k=4$ and electron temperature $T=10^5$ K, consistent with solar wind observations. To reconstruct the distribution functions defined in equation~(\ref{eq1}), approximately 10$^6$ test-particle electrons are followed. The resulting upstream energy spectra shown in the following sections represent the reflected electron populations, and are normalised to the total number of particles injected in the simulation.

\begin{figure*}
	\includegraphics[width=\textwidth]{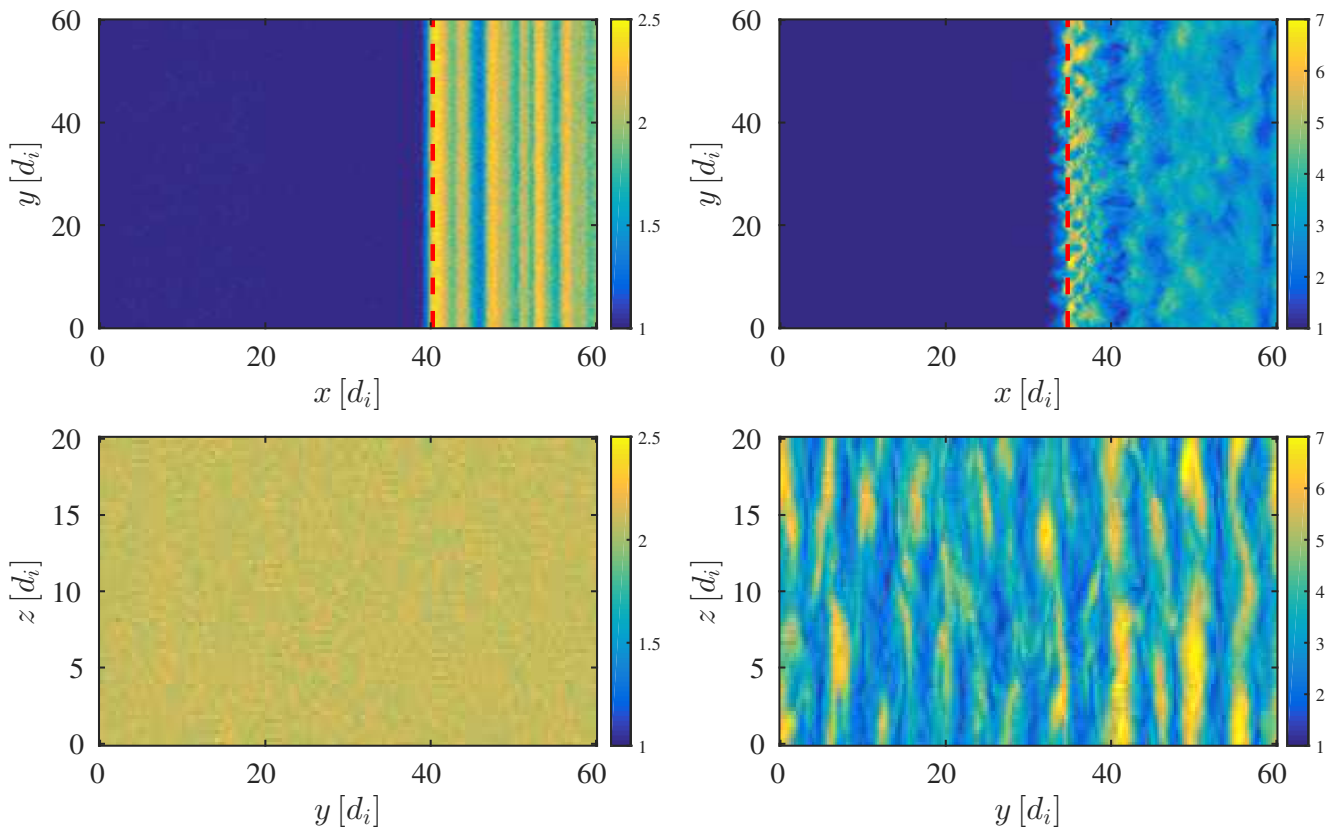}
   	 \caption{Magnetic field intensity plots for low (left) and high (right) Mach number shocks (M = 2.9 and 6.6, 
	 	       respectively). Top panels correspond to cuts of 3D data along $z$ = 10 $d_i$.	 	        The bottom panels correspond to cuts along the red dashed lines in the plots above). In both cases, the upstream $\theta_{Bn}$ is 87$^\circ$.}
    \label{fig:fig1}
\end{figure*}

\section{Results}
\label{sec:results}

\subsection{Shock ramp fluctuations}
\label{sec:shock_fluc}

In order to set the context for the shock structure, we summarize results from 3D hybrid simulations at different Mach numbers. Figure~\ref{fig:fig1} shows the magnetic field amplitude at two shocks with $\theta_{Bn}$ = 87$^\circ$ (with the upstream magnetic field in the $x$-$y$ plane) and M$_A$ = 2.9 and 6.6, respectively. Cuts in the $x$-$y$ and $y$-$z$ simulation planes are shown. The $y$-$z$ cut is taken at the $x$ position indicated with the red dashed line, corresponding to the shock ramp position, and shows the structuring of the shock surface in the field direction and perpendicular to the coplanarity plane. In the low Mach number case the shock has an approximately smooth surface, whereas at high Mach number the shock surface appears to be considerably structured (note that the colour scale ranges are different in the two cases). The high Mach number case shows rippling over a range of wavelengths, primarily in the $y$ direction with wave vectors parallel to $\vec{B}$. This is consistent with the appearance of field-aligned propagating ripples associated with the reflected-gyrating ions seen in the foot of the shock \citep{Winske1988,Lowe2003}, when the shock is supercritical. There is also strong structuring in the $z$ (out-of-coplanarity plane) direction, which is completely absent in 2D simulations in the $x$-$y$ plane. The formation mechanisms for structure seen in the shock surface in 3D hybrid simulations is discussed in \citet{Burgess2016}. When 2D high Mach number shock simulations are performed, with the upstream magnetic field in the simulation plane, the structure is dominated by the field-aligned propagating ripples. The field-aligned ripple structure in 2D simulations can be suppressed by choosing the upstream magnetic field to point out of the simulation plane \citep{Burgess2006}, although in this case, for some upstream parameters, there can be other sources of structuring \citep{Burgess2007}.

\begin{figure*}
	\includegraphics[width=\textwidth]{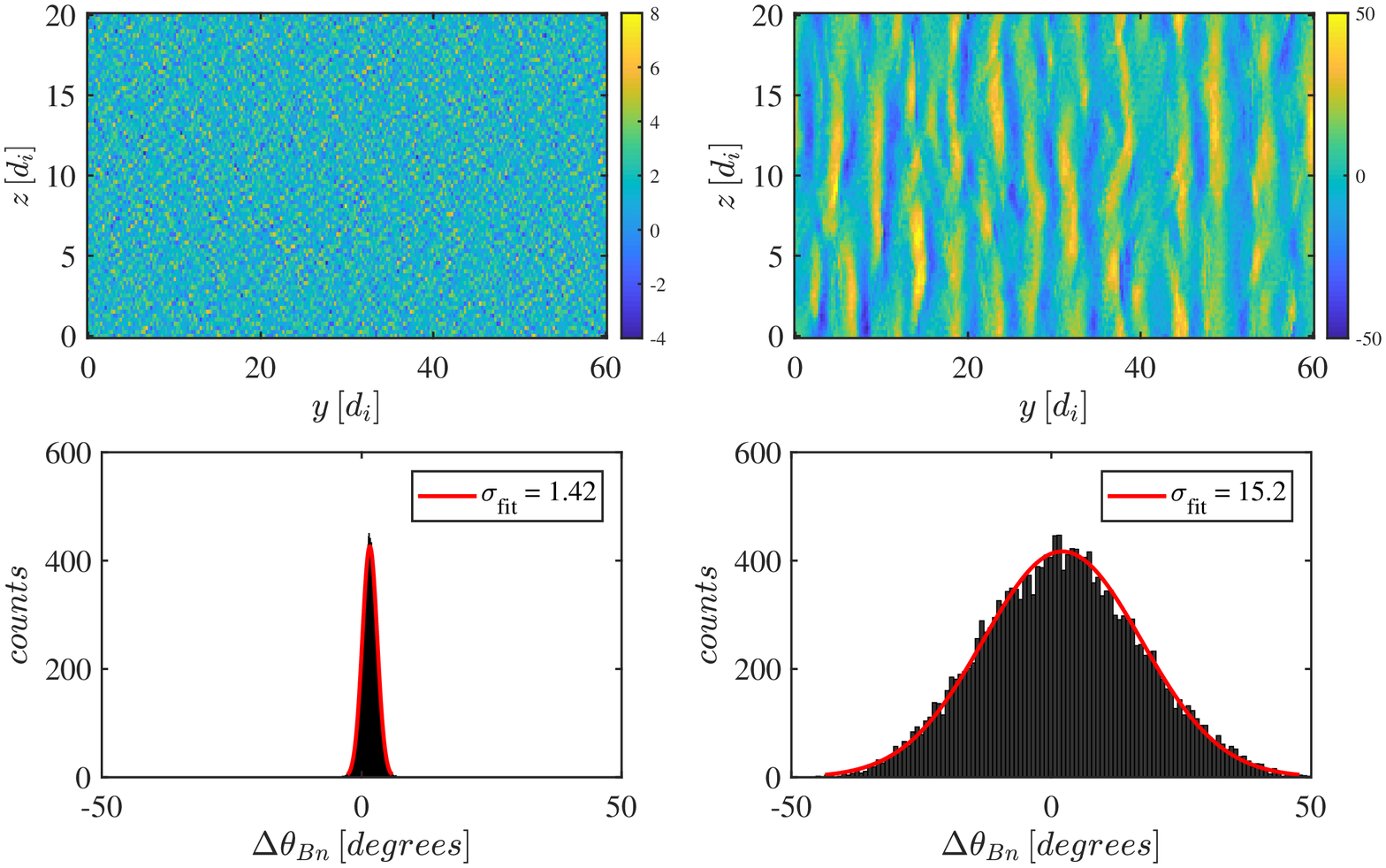}
   	 \caption{Top panels: Map in $z$-$y$ plane of $\Delta\theta_{Bx}$, the deviation of $\theta_{Bx}$ from its nominal upstream value, over the shock front for low (left) and  high (right) Mach number shocks ($M=2.9$ and 6.6, respectively). The $x$ position of the planes shown are as given in Fig.~\ref{fig:fig1}. Bottom panels: Histograms of $\Delta\theta_{Bx}$ around the shock front and respective gaussian fits. }
    \label{fig:fig2}
\end{figure*}

It is interesting to identify potential sources of electron scattering at supercritical shocks, and in particular the local transient changes in geometry that are responsible for departures from adiabatic theory. Figure~\ref{fig:fig2} shows the distribution of the local $\theta_{Bx}$ value over the shock surface in the $y$-$z$ plane for the two shocks shown in Figure~\ref{fig:fig1}. In a small region (1 $d_i$ thick) around the shock the value of $\theta_{Bx}$ is calculated at each cell, using the magnetic field at that position, and the occurrence distribution is accumulated. At low Mach number, there are only small deviations from the nominal upstream value of $\theta_{Bx}$, mostly due to the statistical noise present in the simulation, whereas for the high Mach number case there is a broad range of $\theta_{Bx}$ due to fluctuations. To make these statements more quantitative, a Gaussian distribution is fitted to the histogram of the deviation from nominal value $\Delta\theta_{Bx}$, and then its standard deviation $\sigma$ is a measure of the fluctuation level.
From Figure~\ref{fig:fig2}, $\sigma\approx 1.4^\circ$ for the low Mach number shock front, which is considerably smaller than in the high Mach number case ($\sigma\approx 15.2^\circ$). Note, some of the variation in $\theta_{Bx}$ will be due to the change in the field angle across the shock due to the conservation relations, but since the shock is close to perpendicular this is very small. Electrons, with their small gyroscale and high velocity, will sample the variation of the local $\theta_{Bx}$ within the shock region, giving more favourable conditions for scattering in the high Mach number shock.

\subsection{Electron acceleration: 2D and 3D shock simulations}
\label{sec:2D_3D}

\begin{figure}
	\includegraphics[width=\columnwidth]{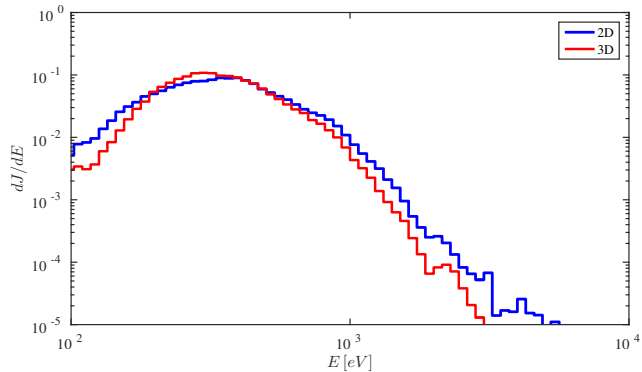}
   	 \caption{Comparison of upstream collected electrons energy spectra in 2D and 3D simulations in low M regime. The Mach number is 2.9 and the upstream $\theta_{Bn}$  is 87$^\circ$.}
    \label{fig:fig3}
\end{figure}

\begin{figure}
	\includegraphics[width=\columnwidth]{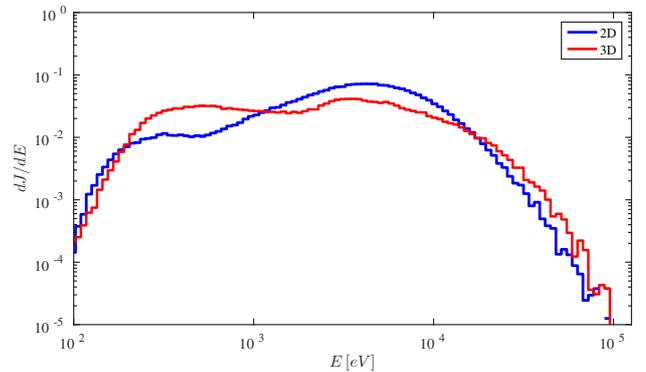}
   	 \caption{Comparison of upstream collected electrons energy spectra in 2D and 3D simulations in high M regime. The Mach number is 6.6 and the upstream $\theta_{Bn}$  is 87$^\circ$.}
    \label{fig:fig4}
\end{figure}

In this section we compare the results of the test particle simulations using upstream electron energy spectra obtained with 2D and 3D hybrid simulation datasets. The spectra are collected upstream of the shock, after all the test-particle electrons have interacted with the shock front; electrons that did not interact with the shock are excluded. The electrons are initialised upstream as a kappa distribution with $\kappa$ = 4, consistent with solar wind observations. In Figures~\ref{fig:fig3} and~\ref{fig:fig4} we show the energy spectra for low and high Mach number shocks with a nominal $\theta_{Bn}$ of 87$^\circ$.
In the low Mach number case, shown in Figure~\ref{fig:fig3}, the resulting distribution from the shock interaction shows only moderate energisation, irrespective of whether the simulation is 2D or 3D. The results in this Mach number regime are consistent with what would be expected from adiabatic theory with a static planar shock. The spectrum in the 2D case shows a slightly higher flux at high energies, but, as discussed in the next section, this is probably an artefact of the two-dimensionality of the shock simulation.

The energy spectra in high Mach number case (Figure~\ref{fig:fig4}) show an important change of character, so that energisation is seen to be much more efficient when the shock has a rippled surface. In particular, the distribution function exhibits a plateau-like feature extending in energy over two decades. This behaviour is a consequence of the additional source of scattering provided by the broadband fluctuations at the shock front, confirming the results obtained previously with 2D simulations \citep{Burgess2006}. Comparing the spectra obtained from 2D and 3D datasets, overall there are not strong differences, and in particular the maximum energisation obtained for the electrons in both cases is approximately the same. The falloff of the 3D spectrum is shallower than the 2D one. This behaviour is attributed to the fact that in the 3D hybrid simulations, the structuring of the shock front is fully resolved in the $z$ direction, introducing an additional source of scattering through the fluctuations present at the shock front. 

\subsection{Mach number and $\theta_{Bn}$ dependence}
\label{sec:3D_spectra}

\begin{figure}
	\includegraphics[width=\columnwidth]{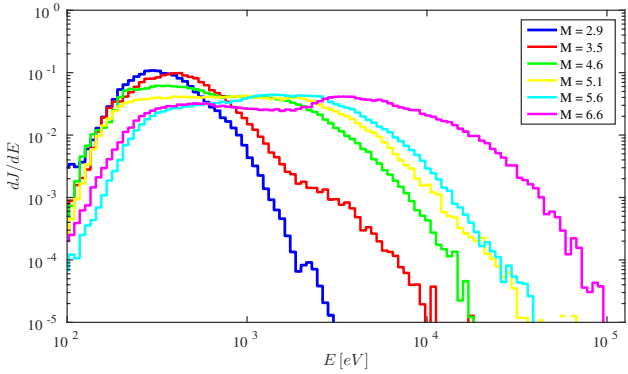}
   	 \caption{Comparison of final upstream electron energy spectra for 3D shocks with different Mach numbers. In all the cases the upstream $\theta_{Bx}$ is 87$^\circ$.}
    \label{fig:fig5}
\end{figure}

Throughout this section, results using 3D hybrid shock and test particle simulations are shown. Figure~\ref{fig:fig5} shows the final upstream electron energy spectra for $\theta_{Bn}$ = 87$^\circ$ as the Mach number varies from 2.9 to 6.6. The appearance of the plateau-like feature in the spectra is controlled by the shock Mach number and, hence, by the presence of shock surface fluctuations. The lowest Mach number case (M = 2.9), as shown previously (Figure~\ref{fig:fig3}), is compatible with the theory of adiabatic reflection. At slightly higher Mach number (M=3.5), the shock front starts to exhibit a moderate level of fluctuations, resulting in a longer tail in the spectrum and leading to a maximum  electron energisation of 10 keV. At higher Mach numbers, the shock front exhibits fully developed surface fluctuations, leading to higher maximum energies and to the flattening of the spectrum as discussed above.

\begin{figure}
	\includegraphics[width=\columnwidth]{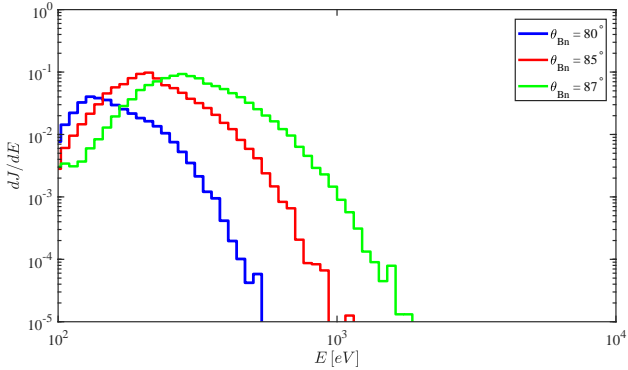}
   	 \caption{Comparison of final upstream electron energy spectra for three 3D shocks with different $\theta_{Bx}$. In all cases, the Mach number of the shock is 2.9.}
    \label{fig:fig6}
\end{figure}
\begin{figure}
	\includegraphics[width=\columnwidth]{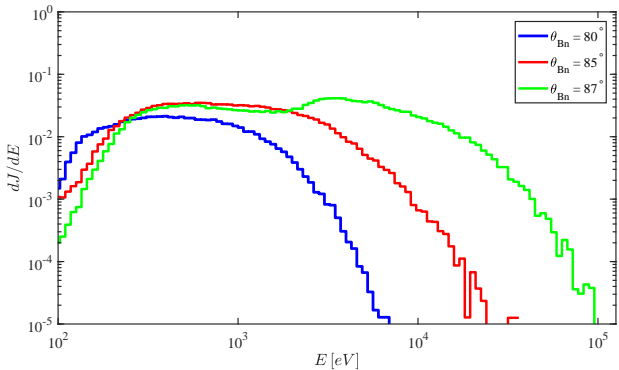}
   	 \caption{Comparison of final upstream electron energy spectra for three different $\theta_{Bx}$ cases. The Mach number of the shock is M = 6.6 .}
    \label{fig:fig7}
 \end{figure}

Adiabatic reflection acceleration theory predicts that effective electron energisation is possible only for angles $\theta_{Bn}$ very close to perpendicular. Simulations in 3D were performed for a range of geometries between 80$^\circ$ and 87$^\circ$ for both  low and high Mach number regimes to gain information about the role shock surface fluctuations can play in broadening the range of geometries for which effective electron acceleration is observed. Figure~\ref{fig:fig6} shows upstream electron spectra for simulations in the low Mach number regime for $\theta_{Bn}= 80^\circ$, $85^\circ$ and $87^\circ$. It is clearly seen that increasing $\theta_{Bn}$ produces a more efficient energisation of the electron population. Again, this is consistent with the predictions from adiabatic theory. For the $\theta_{Bn} = 80^{\circ}$ case, the energisation is rather weak, with a maximum electron energisation less than 1 keV.

Figure~\ref{fig:fig7} shows the same set of $\theta_{Bn}$ values, but for the high Mach number (M = 6.6) regime. An important increase in energisation going from smaller to larger $\theta_{Bn}$ values is still observed, but a significant flattening of the spectra is found, even in the case $\theta_{Bn}=80^\circ$. This shows that the presence of shock surface fluctuations can lead to significant energisation even for more oblique geometries. 

Summarising, the enhancement of electron acceleration due to shock surface fluctuations is evident when transitioning from small to high Mach numbers (and hence from sub to super critical regimes), as shown in Figure~\ref{fig:fig5}. Furthermore, the presence of shock surface fluctuations broadens the range of $\theta_{Bn}$ at which efficient electron acceleration is found (Figures~\ref{fig:fig6} and \ref{fig:fig7}).

\section{Electron dynamics}
\label{sec:discussion}

\begin{figure}
	\includegraphics[width=\columnwidth]{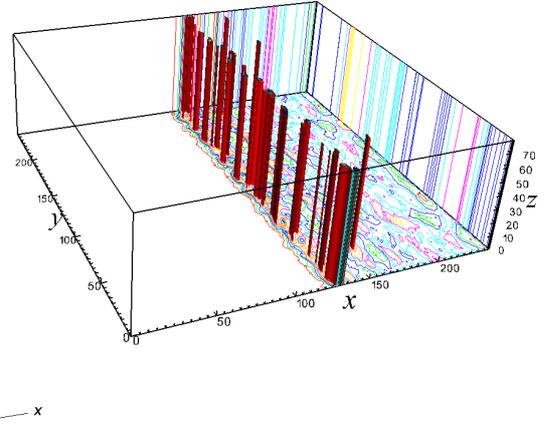}
   	 \caption{2D contours and 3D isocontours of magnetic field magnitude for a 3D hybrid shock simulation (M = 6.6).}
    \label{fig:fig8}
\end{figure}

\begin{figure}
	\includegraphics[width=\columnwidth]{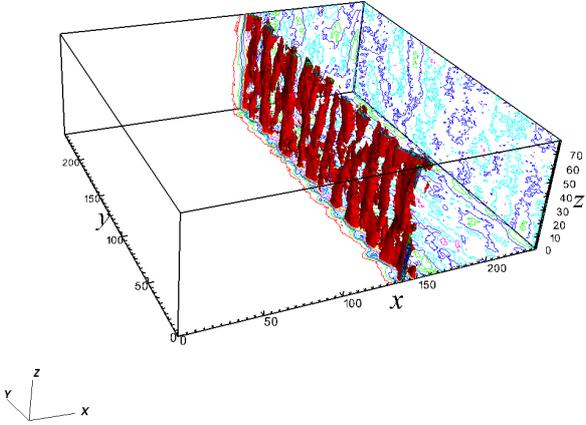}
   	 \caption{2D contours and 3D isocontours of magnetic field magnitude for a 2D hybrid shock simulation (M = 6.6), shown as a 3D dataset. The 2D simulation output is replicated along the $z$ direction.}
    \label{fig:fig9}
\end{figure}

In order to understand the nature of the ripples experienced by the electrons, Figure~\ref{fig:fig8} shows a plot of isosurfaces of the magnetic field magnitude for a high Mach number, M$_A$ = 6.6, $\theta_{Bn}$ = 87$^\circ$ 3D shock simulation. The field lines are principally in the $y$ direction, and as electrons move along the field lines the structure within the shock is able to add additional scattering through broadband fluctuations. Additionally, the elongated structures in the $z$ direction can act as electron traps which boost energization, since the electron energy gain is mainly via drift motion along the $E_z$ electric field. Figure~\ref{fig:fig9} shows a similar plot for a 2D shock simulation with the same parameters, with invariance in the $z$ dimension. This figure merely illustrates the coherence of the ripples in the $z$ dimension, produced by the reduced dimensionality of the simulation, which can make them a more efficient trap for electrons, in comparison to the 3D case. Some evidence for this may be, for the 2D case, the increase in the energy spectrum before its eventual high energy fall off (Figure \ref{fig:fig4}).
 

\begin{figure}
	\includegraphics[width=\columnwidth]{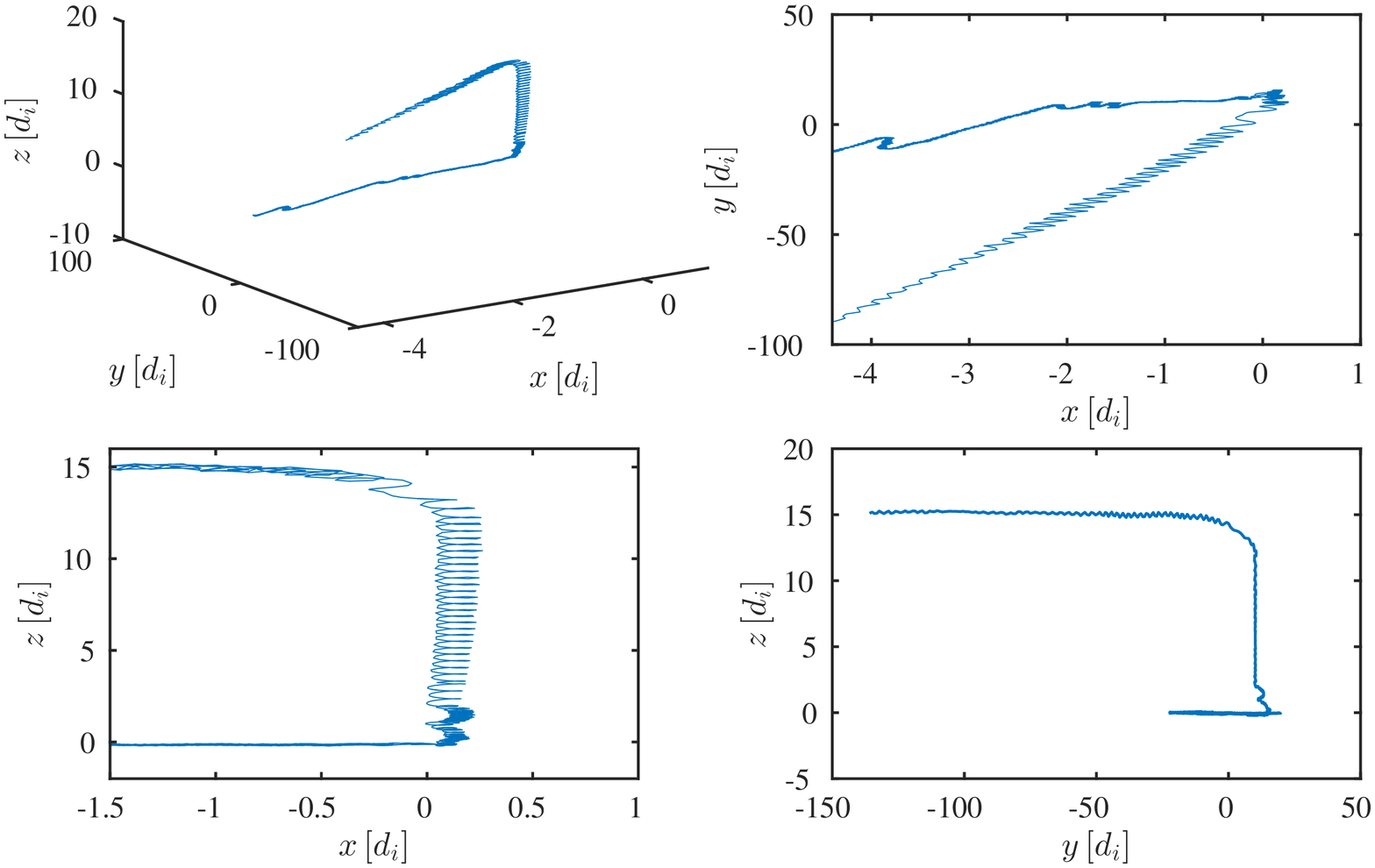}
   	 \caption{Particle trajectory in 3D and with its 2D projections for an electron interacting with a 2D, 87$^\circ$, M = 2.9 shock.}
    \label{fig:fig10}
\end{figure} 
 \begin{figure}
	\includegraphics[width=\columnwidth]{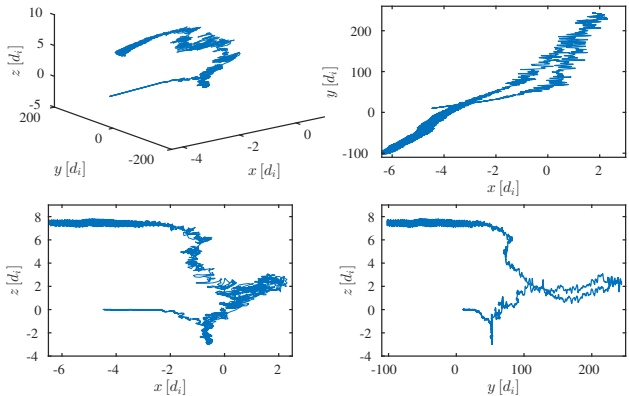}
   	 \caption{Particle trajectory in 3D and with its 2D projections for an electron interacting with a 3D, 87$^\circ$ M=6.6 shock.}
    \label{fig:fig11}
 \end{figure}

To corroborate this interpretation, two typical trajectories of electrons accelerated at the shock front are presented. Figure~\ref{fig:fig10} shows the trajectory of a test-particle electron interacting with a 2D, low Mach number shock. The particle has an initial energy of 100 eV, and a final energy of 2 $\times$ 10$^3$ eV. The zero on the $x$-axis corresponds to the nominal shock position, so the trajectory is plotted in the normal incidence shock frame. At the shock ramp, the particle travels for a very long distance along $z$, with little motion in the $x$-$y$ plane, demonstrating artificial enhanced trapping at the shock front. Here artificial means that it is an artefact of the reduced dimensionality of the the shock simulation. From examination of many trajectories in the low Mach number regime, the subset of artificially trapped electrons was identified. It was found that they systematically reach higher energies than the ones that are reflected at the shock front with a motion overall contained in the $x$-$y$ plane. The latter particles show lower energisation, in line with the prediction of adiabatic reflection. The final energies of artificially trapped electrons typically belong the tail of the resulting upstream distributions shown in Figure~\ref{fig:fig3}, explaining the slightly higher fluxes found in spectra obtained from 2D simulations compared to those obtained by 3D ones.

As comparison, Figure~\ref{fig:fig11} shows the trajectory of an electron in a 3D, high Mach number shock. This particle, also initialised with an energy of 100 eV, exhibits very efficient energisation (the final energy is 3.4 $\times$ 10$^4$ eV), happening mostly when it travels a long way in the $z$ direction (see the lower right plot). The $z$ motion consists of several phases of drift and energy gain, as if it is interacting with more than one ripple throughout the reflection process. This confirms the scenario of the shock ripples acting as an extra scattering source and trap at the same time, causing electrons to undergo stochastic acceleration while interacting with the ripples. A relatively large number of trajectories exhibit this type of behaviour. It is also interesting to note that these particles penetrate relatively far downstream of the shock front, before eventually being scattered upstream again. 

\begin{figure}
	\includegraphics[width=\columnwidth]{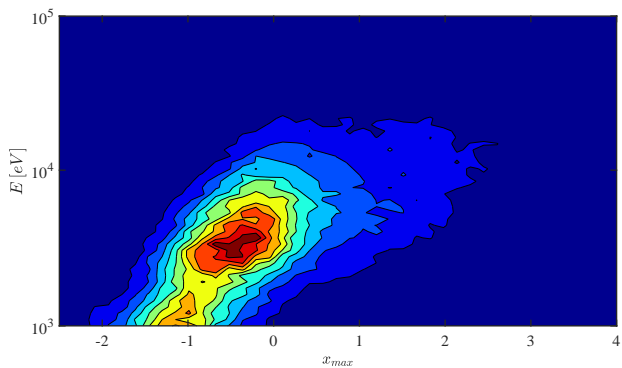}
   	 \caption{Maximum particle position relative to shock as a function of electron energisation for an ensemble of electrons interacting with a 3D, 87$^\circ$ high Mach number shock. The initial electron distribution is a 100 eV monoenergetic shell.}
    \label{fig:fig12}
\end{figure}

To demonstrate the role of scattering immediately downstream of the shock ramp Figure~\ref{fig:fig12} shows the distribution of maximum $x$ position and final particle energy of test-particle electrons that are subsequently collected upstream. The $x$ position is shown with respect to the nominal shock position. Hence, the electrons are going to experience their first interaction with the shock in a region surrounding the nominal $x$ = 0 that is approximately 1 $d_i$ thick. All the electrons in the high energy region of the upstream electron spectrum (energy > 10$^3$ eV) have been selected for this distribution. The maximum particle position, $x_{max}$, is the maximum position reached along $x$ by an electron from its release to the upstream collection. This distribution is for the case of an initial single monoenergetic shell of 100 eV electrons. It can be seen that the more energetic particles are the ones which reach a maximum position deeper into the shock, and so spend longer interacting with it. It is natural that these particles are also those which have suffered additional scattering associated with the rippled shock structure.

\section{Conclusions}
\label{sec:conclusions}

We have performed 2D and 3D hybrid simulations of quasi-perpendicular collisionless shocks and used the test particle method to study electron acceleration starting from suprathermal energies. It is the first time that the full, three-dimensional structure of the shock has been considered in the framework of hybrid plasma and test-particle simulations. The strongest signatures of electron acceleration were found for shocks with a shock normal geometry close to perpendicular. The difference between electron acceleration at subcritical (low Mach number) and supercritical (high Mach number) shocks was addressed with both 2D and 3D simulations. There is a clear difference between these two cases. At low Mach number only moderate energisation is found, consistent with single interaction, coherent reflection models based on adiabatic motion \citep{Leroy1984,Wu1984}. In the supercritical regime enhanced energization is found, which can be explained by considering the small scale structures and fluctuations in the shock ramp \citep[e.g.,][]{Burgess2006,Guo2010}, which produce additional scattering.

Comparing 2D and 3D simulations, in 2D we find signatures of enhanced acceleration due to particle trapping in fluctuations, which is an artefact of the reduced dimensionality. In the subcritical regime this leads to slightly higher electron energy gains, and the mechanism appears to be important also for the supercritical regime, accentuating the plateau feature observed in upstream energy spectra. 

When the full three dimensional structure of the shock front is resolved, slightly higher final electron energies are obtained in the supercritical regime.  We believe that this is due to the possibility that the electrons can interact with several surface fluctuations throughout the reflection process, thus being retained in the shock transition layer for longer times where they experience the electric field parallel to the shock surface ($E_z$) responsible for their acceleration. This scenario is corroborated by the analysis of Figure~\ref{fig:fig12}, which illustrates that electrons that penetrate deepest into the shock layer before reflection have the largest final energies. Therefore, any process which tends to retain the electrons at the shock front increases their energy gain. Although the current simulations rely on ion scale fluctuations, a similar scenario has been proposed based on full kinetic PIC simulations \citep{Amano2009}.

In the supercritical regime, efficient electron acceleration was also found at more oblique shock geometries, down to $\theta_{Bn}$ = 80$^\circ$ (Figure~\ref{fig:fig7}). It is important to consider that the key ingredient for efficient electron acceleration is the local value of $\theta_{Bn}$, rather than the average one. In this respect, the shock rippling causes local changes in the shock geometry at various spatial scales (see Figures~\ref{fig:fig1} and \ref{fig:fig2}), so when an electron approaches the shock transition, it has a certain possibility to interact with a local part of the shock with temporarily perpendicular geometry, and then later a less perpendicular region allowing upstream escape. This phenomenon leads, statistically, to higher energisation when rippling is present at the shock front. 
 
A number of restrictions in this study should be noted. The shock simulations do not include any large scale non-stationarity (such as shock reformation), and the shock fluctuations are self-generated without accounting for any upstream turbulence, which can be important \citep{Guo2010,Guo2015}. The relative roles of these two mechanisms remains a topic for future work. As initial electron distributions kappa functions have been used, motivated by solar wind observations \citep[e.g.,][]{Maksimovic2005}, but similar results have been found using different initial distributions (e.g., Maxwellian). However, electron distribution functions and their features need more observational work for a full characterization in the solar wind \citep[e.g.,][]{Graham2017}. A major issue with this study is the use of a reduced plasma model, such as the hybrid model, for the shock structure fields. The justification for using a hybrid simulation combined with test particles is based on starting from electron superthermal energies, assuming that the electron scale structure (which is absent in the hybrid approximation) does not play an important role for the superthermal electrons. This allows longer time periods to be simulated, so the effects of ion scale structure can be shown to be important. However, the results should be confirmed by fully kinetic simulations, although they have their own limitations such as small spatial extents and unrealistic mass ratio \citep[e.g.,][]{Krasno2013}. From our results we see that electron acceleration is efficient when there is a combination of magnetic mirror reflection and scattering within the shock gradient. There is evidence from fully kinetic PIC simulations that a similar process operates for the thermal electrons \cite{Amano2009}, although the fluctuations producing the scattering are different in nature (lower hybrid rather than ion scale ripples). In either case, depending on initial energy, the importance of scattering is key for efficient acceleration. The additional point made by this work is that the presence of ion scale ripples as a source of scattering requires that the Mach number should be supercritical. Finally, as for all simulation work, validation for this picture is required by observational means. 
 
It is important to remark that the critical Mach number, at which shocks develop surface fluctuations, is relatively low ($M_c \approx 3.5$). Solar wind observations show that, at 1 AU, the range of observed shock parameters covers both sub- and supercritical regimes, so it is possible, in principle, to test the results of analytical and numerical studies. However, further observational work is required towards a deeper understanding of electron acceleration: in particular in order to investigate the role of upstream solar wind turbulence, which is expected to enhance the electron energisation. Another challenging topic is to look at how sub- and supercritical shock regimes can effect electron injection into other mechanisms leading to higher energies and observed in large scale  astrophysical environments, such as giant radio relics in the intracluster medium \citep[e.g.,][]{Brunetti2016,Kang2017}, and this will be the object of future investigations.

\section*{Acknowledgements}
DT acknowledges support of a studentship funded by the Perren Fund of the University of London. This research was supported by the UK Science and Technology Facilities Council (STFC) grant ST/P000622/1 (D. Burgess). This research utilised Queen Mary's Apocrita HPC facility, supported by QMUL Research-IT, http://doi.org/10.5281/zenodo.438045.




\bibliographystyle{mnras}
\bibliography{mybib} 

\bsp	
\label{lastpage}
\end{document}